%% file: main.tex
\PassOptionsToPackage{dvipsnames}{xcolor}
\documentclass[conference]{IEEEtran}
\IEEEoverridecommandlockouts
\usepackage{cite}
\usepackage{amsmath,amssymb,amsfonts}
\usepackage{algorithm}
\usepackage{algpseudocode}
\usepackage{hyperref} 
\usepackage{bookmark}
\usepackage{graphicx}
\usepackage{textcomp}
\usepackage{xcolor}
\usepackage{svg}
\usepackage{enumitem}
\usepackage{multirow}
\usepackage{listings}
\usepackage{lstautogobble}
\usepackage{subcaption}
\usepackage{booktabs}
\usepackage{tikz}
\usepackage{stfloats}
\usepackage{fancyhdr}
\usepackage{sc26repro}

\usepackage{tabularx,booktabs}

\setlist{nolistsep}

\usepackage{threeparttable}

\def\BibTeX{{\rm B\kern-.05em{\sc i\kern-.025em b}\kern-.08em
    T\kern-.1667em\lower.7ex\hbox{E}\kern-.125emX}}

\newcommand{\didem}[1] {{\color{blue}{Didem}}: {\color{red}{#1}}}
\newcommand{\sinan}[1] {{\color{blue}{Sinan}}: {\color{red}{#1}}}

\newcommand{\james}[1] {{\color{blue}{James}}: {\color{red}{#1}}}

\makeatletter
\DeclareUrlCommand\ULurl@@{%
  \def\UrlLeft{\uline\bgroup}%
  \def\UrlRight{\egroup}}
\def\ULurl@#1{\hyper@linkurl{\ULurl@@{#1}}{#1}}
\DeclareRobustCommand*\ULurl{\hyper@normalise\ULurl@}
\makeatother

\definecolor{bluekeywords}{rgb}{0.13, 0.13, 1}
\definecolor{greencomments}{rgb}{0, 0.5, 0}
\definecolor{redstrings}{rgb}{0.9, 0, 0}
\definecolor{graynumbers}{rgb}{0.5, 0.5, 0.5}
\definecolor{Teal}{RGB}{0, 128, 128}

\definecolor{background_ok}{rgb}{0.95, 0.95, 0.95}

\usepackage[listings,skins]{tcolorbox}
\def\BibTeX{{\rm B\kern-.05em{\sc i\kern-.025em b}\kern-.08em
    T\kern-.1667em\lower.7ex\hbox{E}\kern-.125emX}}

\usepackage[switch]{lineno}
   
\usepackage{tcolorbox}

\begin{document}

\title{Triple Overlap in Pipelined Krylov Methods on Multi-GPU Systems}

\makeatletter
\newcommand{\linebreakand}{%
  \end{@IEEEauthorhalign}
  \hfill\mbox{}\par
  \mbox{}\hfill\begin{@IEEEauthorhalign}
}
\makeatother

\author{
\IEEEauthorblockN{Sinan Ekmekçibaşı}
\IEEEauthorblockA{\textit{Koç University} \\
Istanbul, Turkey \\
sekmekcibasi23@ku.edu.tr}
\and
\IEEEauthorblockN{James D. Trotter}
\IEEEauthorblockA{\textit{Simula Research Laboratory} \\
Oslo, Norway \\
james@simula.no}
\and
\IEEEauthorblockN{Johannes Langguth}
\IEEEauthorblockA{\textit{Simula Research Laboratory}, Oslo, Norway \\
\textit{University of Bergen}, Bergen, Norway \\
langguth@simula.no}
\linebreakand
\IEEEauthorblockN{Xing Cai}
\IEEEauthorblockA{\textit{University of Oslo}, Oslo, Norway \\
\textit{Simula Research Laboratory}, Oslo, Norway \\
xingca@ifi.uio.no}
\and
\IEEEauthorblockN{Didem Unat}
\IEEEauthorblockA{\textit{Koç University} \\
Istanbul, Turkey \\
dunat@ku.edu.tr}
}

\maketitle

\begin{abstract}

We introduce a novel triple-overlap implementation of the pipelined preconditioned conjugate gradient (Pipe-PCG) and pipelined stabilized biconjugate gradient (Pipe-BiCGStab) solvers, methods widely used in scientific applications. By leveraging a collective communication library, our approach extends conventional computation–communication overlap to concurrently execute collective operations, irregular P2P communication, and compute kernels, enabling computation–communication–communication overlap across multiple GPUs.
We evaluate the design on the MareNostrum 5 (H100) and Leonardo (A100) supercomputers, comparing against PETSc and the state-of-the-art two-stream NCCL implementation using SuiteSparse matrices and three massive instances scaled up to 512 GPUs. The three-stream design outperforms the two-stream baseline on every instance, with the margin widening from 2–6\% at 4 GPUs to 18–47\% at 64 GPUs. At 64 GPUs, the geometric-mean speedup over the two-stream baseline is 1.33× on MareNostrum 5 and 1.23× on Leonardo for Pipe-PCG, and reaches 1.30× for Pipe-BiCGStab. In large scale experiments, we observed a speedup up to 35\% in Pipe-PCG and up to 33\% in Pipe-BiCGStab compared to two-stream design. For PETSc, the margin is even wider, up to 3.27x speed-up in both Pipe-PCG and Pipe-BiCGStab. Related code for the proposed method is available online: \hyperlink{https://github.com/ParCoreLab/aCG}{https://github.com/ParCoreLab/aCG}


\end{abstract}


\input{sections/introduction}

\input{sections/background}
\input{sections/implementation}

\input{sections/experiments}

\input{sections/related_works}

\section{Conclusion}
We have introduced the three-stream technique for overlapping two different types of communication - high volume, bandwidth limited P2P messages and low volume latency limited collectives. The idea can be implemented efficiently on GPU systems using three separate streams for concurrency, making use of modern GPU communication libraries such as NCCL or RCCL which can handle communication in streams.
Our experimental results on Pipe-PCG and Pipe-BICGStab indicate that the technique works as intended, yielding substantial performance improvements when scaling to a large number of GPUs. 
In future work, we intend to explore other applications and plan to implement the three-stream technique for AMD GPUs.

\section*{Acknowledgments}
Authors from Koç University have received funding from the European Research Council (ERC) under the European Union’s Horizon 2020 research and innovation programme (grant agreement No 949587).

We acknowledge EuroHPC Joint Undertaking for awarding access to the MareNostrum5 supercomputer in Spain and the Leonardo supercomputer in Italy (Project IDs: EHPC-DEV-2024D10-091, EUHPC\_D21\_118).

Various AI-based tools, including Claude Opus, OpenAI’s ChatGPT, and Google’s Gemini Pro, were used throughout the preparation of this manuscript to assist with language editing and LaTeX formatting. All scientific content, data, analyses, and conclusions were independently reviewed and verified by the authors.

\bibliographystyle{IEEEtran} 
\bibliography{references}

\input{sections/AD}

\vspace{12pt}
\color{red}

\end{document}

%% file: sections/introduction.tex
\section{Introduction}
\label{sec:intro}

Overlapping communication with computation is a well-established approach to reducing communication cost. A common example is the use of non-blocking \textit{send} and/or \textit{receive} operations to initiate individual message transfers, typically between neighboring processes to exchange halo data \cite{halo-exchange-lib, Comm-Comp-Overlap}. When these messages make progress in the background, computation can proceed concurrently. 
 This strategy is effective in both multi-CPU and multi-GPU settings. On CPUs, message progression relies on dedicated communication threads \cite{langguth2015scalable} or advanced MPI implementations \cite{ruhela2018efficient}. On GPUs, overlap is typically realized with two concurrent streams, one for communication and one for computation \cite{trotter2025cpu, uniconn}.

The conjugate gradient (CG) method \cite{CG-method}, its preconditioned (PCG) variants and pipelined stabilized biconjugate gradient (Pipe-BiCGStab) methods are canonical Krylov subspace solvers for sparse linear systems, which makes them a central building block in large-scale scientific applications. Prior work on communication overlap in CG has largely treated its two communication types separately: global collective communication needed for parallelizing dot products and point-to-point (P2P) halo exchange in distributed sparse matrix-vector multiplications (SpMV). The literature on pipelined Krylov methods addresses the former to lower the global synchronization overhead, by overlapping non-blocking reductions with the SpMV and preconditioner operations, and this idea is further extended through deeper pipelines or fewer reductions per iteration \cite{ghysels2014hiding, deeppipelineCG, tiwari-DistributedMemCG, sanan16}.
The pipelined CG algorithm first presented in~\cite{ghysels2014hiding} is theoretically capable of overlapping both collective and P2P communication with computation. The authors of that work, however, focused only on hiding collective operations.

A separate line of work targets overlapping halo/P2P data exchanges with computation
in iterative solvers and multigrid methods, primarily through scheduling and implementation techniques, such as specialized OpenMP threads \cite{Nakajima2022OpenMP, Nakajima2025CCOverlapMG}.
On GPUs, recent efforts improve the communication path through multi-GPU SpMV overlap, persistent kernels, and CPU- or GPU-initiated communication mechanisms \cite{PasquaCG, cpu-free, trotter2025cpu,ma2025efficient}. However, none of these works jointly schedule collective communication, halo exchange, and computation as concurrent components of a single design.

In this work, we show that pipelined Krylov subspace solvers such as Pipe-PCG and Pipe-BiCGStab offer richer overlap opportunities than previously exploited. Beyond the traditional communication-computation overlap, they naturally allow \emph{communication-communication} overlap, where collective reductions and P2P halo exchanges proceed concurrently alongside computation. To exploit this, we propose a novel \emph{three-stream} implementation of Pipe-PCG and Pipe-BiCGStab that overlaps three components: collective reduction, halo exchange, and SpMV. Our design builds on the NVIDIA Collective Communication Library (NCCL)~\cite{NCCL}, which provides native stream support for GPU communication.

Our contributions are as follows:

\begin{itemize}
    \item We propose a three-stream overlap strategy for pipelined preconditioned CG and pipelined BiCGStab solvers that concurrently overlaps collective communication, P2P communication, and computation.




    \item We evaluate our three-stream strategy against PETSc and a two-stream baseline \cite{trotter2025cpu}. Over this baseline, we achieve up to 35\% (Leonardo, 256 GPUs) and 44\% (MareNostrum~5, 64 GPUs) speedups for Pipe-PCG, with similar peak gains of 33\% (MareNostrum5, 256 GPUs) and 42\% (MareNostrum~5, 64 GPUs) for Pipe-BiCGStab compared to two-stream baseline.
    
\end{itemize}

Beyond Pipe-PCG and Pipe-BiCGStab, the three-stream design can be generalized to any application combining high-bandwidth local data exchange with high-latency global coordination (e.g., convergence checks or global updates).

%% file: sections/background.tex
\section{Background and Motivation}
\label{sec:background}

\subsection{Pipelined Preconditioned CG (Pipe-PCG)} 
Pipe-PCG methods \cite{ghysels2014hiding} were developed to address the severe communication bottlenecks that limit the large-scale parallel scalability of standard PCG, 
where each iteration requires two separate dot products. In a distributed memory environment, each dot product forces a global synchronization event. These frequent synchronizations expose network latency and cause some processes to sit idle, creating a start-and-stop execution model that severely inhibits parallel efficiency \cite{ghysels2014hiding}. 


Pipe-PCG directly targets this synchronization bottleneck by mathematically restructuring the solver to group the two dot products together, reducing the global synchronization to a single non-blocking phase per iteration. More importantly, this synchronization can be delayed until the preconditioner and SpMV are done.
The Pipe-PCG implementation relies on the same fundamental
compute kernels as the standard PCG method: vector additions (AXPY), dot products (DOT), 
and SpMV and a preconditioner. However, because Pipe-PCG 
restructures the algorithm, it introduces 
additional AXPY operations. Specifically, a single iteration of Pipe-PCG performs two DOT products combined, 
eight AXPY operations, one SpMV operation, and a preconditioning operation. 
The main computational difference from standard PCG is the use of eight AXPYs 
instead of three. Although these extra AXPYs increase the overall 
computational load, the benefit from reduced global synchronization 
amortizes this cost during large-scale executions~\cite{ghysels2014hiding}.

\subsection{Pipelined Stablizied Biconjugate Gradient (Pipe-BICGSTAB)}

Standard BICGSTAB suffers from even greater latency exposure than PCG, which requires
three separate global reduction phases to compute four dot products per iteration.
When scaled across distributed-memory architectures, these repeated synchronization
points expose inter-GPU communication latency and force processors into idle wait
states, which inhibits the parallel performance at large scale.

To overcome these communication stalls, Cools and Vanroose~\cite{cools2017pipelined}
introduced Pipe-BICGStab, which restructes the recurrences. By incorporating auxiliary vectors, the algorithm successfully fuses
the dot products into just two global reduction phases. Moreover, these reductions
are formulated as non-blocking collectives that execute concurrently with the
method's two SpMVs, effectively hiding the
network latency behind useful computation.

While using the same core kernels (AXPY, DOT, SpMV), this reformulation increases local computational intensity. Each Pipe-BICGStab iteration executes six dot products (in two collective phases), eight AXPYs, and two SpMVs. Compared to standard BICGStab, it requires four extra AXPYs and expands vector storage from seven to eleven arrays. Despite increased resource demands, hiding inter-GPU communication behind computation yields substantial performance gains on massively parallel systems~\cite{cools2017pipelined}.


\subsection{Beyond Traditional Computation-Communication Overlap}


Since these pipelined Krylov subspace solvers offer more overlap opportunities beyond the conventional computation–communication model, this observation raises the question of whether multiple communication types can be overlapped for greater performance gains.

To answer this question, we first collected some results from \textit{nccl-tests} and 
\textit{BabelStream} microbenchmarks \cite {deakin2018evaluating} to give an indication of relevant computation and communication costs.
Specifically, on a system with 64 GPUs, 
nccl-test's \texttt{allreduce} measurements show that it takes about 45.4\,µs to perform an
\textit{allreduce} operation for two double-precision floating-point numbers, which is required to complete 
the global dot products during a Pipe-PCG iteration. The benchmark also shows that a single P2P message takes approximately the same amount of time when the message size is between 256\,KiB and 512\,KiB. 
Finally, based on \textit{BabelStream}, simple computations such as AXPY and DOT can process about 68.6\,MB of data (corresponding to double arrays with $\sim$8.6 million 
entries) in the same amount of time, i.e., 45\,µs.  This gives a rough idea of the amount of P2P 
communication or computation that will be needed to fully hide the collective (\textit{allreduce}) 
communication that occurs in Pipe-PCG.



Furthermore, we developed a custom three-stream microbenchmark that concurrently executes P2P communication, collective communication, and computation, to see the overlap effect. Specifically, the benchmark consists of an \textit{allreduce} collective on two double-precision scalars (16 bytes), a \textit{send/recv} operation forming a simple ring topology with a payload of 32,000 doubles (~256 KiB), and a DAXPY compute kernel processing 64 MB of data. The baseline for comparison is a traditional two-stream overlap strategy: a compute stream containing the \textit{allreduce} and the DAXPY kernel, alongside a communication stream dedicated to P2P transfers, i.e., DAXPY is overlapped with P2P transfers but the \textit{allreduce} collective is executed without any overlap. 
To mitigate the impact of noise and outlier measurements, we executed both microbenchmarks for $10^5$ iterations. 

We conducted experiments on a single node with four GPUs, comparing the total execution time of the traditional two stream communication-computation overlap with that of our proposed three-way overlap. The three-stream implementation outperformed the two-stream baseline by approximately 5\%. We then inspected the execution using the NVIDIA Nsight Systems profiler. 
The profiler trace demonstrates clear overlap among all three streams: \textit{compute}, \textit{P2P}, and \textit{allreduce}.

%% file: sections/implementation.tex
\section{Implementation}
\label{sec:implementation}

\input{sections/codes/three-stream-code}

This section describes the two-stream and three-stream Pipe-PCG and Pipe-BICGSTAB implementations in detail. We use the two-stream design as the baseline because it achieves state-of-the-art performance among CG implementations~\cite{trotter2025cpu}. 

We implement both the two- and three-stream designs for NVIDIA GPUs using NCCL~\cite{NCCL} for communication, which natively supports GPU streams for asynchronous and overlapping GPU operations. Previous work \cite{trotter2025cpu, desensi2024} showed that on supercomputers NCCL exhibits comparable performance to GPU-aware MPI. 


\subsection{Multi-GPU Pipelined PCG and Pipelined BICGSTAB}

Executing Pipe-PCG and Pipe-BICGStab in a distributed-memory parallel setting requires partitioning the sparse
matrix and its associated vectors. We utilize a row-wise partitioning strategy to 
divide the sparse matrix $A$, distributing the assigned rows and corresponding 
vector segments to each rank. Following this distribution, each GPU becomes 
responsible for its specific partition, allowing the rest of the computations to 
take place locally within each GPU's memory.

\subsubsection{Computation}
In GPU implementations of CG methods, the best practice \cite{mills2025petsc} is to offload the compute
kernels to the GPU and handle intermediate steps such as convergence checks and scalar calculations
on the CPU. For the aforementioned compute kernels, we rely on vendor-optimized BLAS libraries, using
cuBLAS for local dot product and cuSPARSE for SpMV, which employs merge-based SpMV kernels~\cite{merrill2016merge}.


Although vendor libraries offer easier implementation in AXPY operations, going back and forth between host and device to launch the kernels and move scalars between host and device after their calculation may introduce an additional overhead as pointed out by Trotter et al.~\cite{trotter2025cpu}. Therefore, we fused the eight AXPY operations in a single compute kernel together with calculations of $\alpha$ and $\beta$ scalars on the device to mitigate this inefficiency in Pipe-PCG. A similar fusion approach is also applied in Pipe-BICGSTAB, but it is structured differently due to its algorithmic difference. The AXPY operations are grouped into three distinct fused kernels as it can be seen in Figure~\ref{fig:2-3stream}-b, while the scalar computations are performed in a separate CUDA kernel rather than being fused with the AXPYs.

For the preconditioning step in Pipe-PCG, we selected the Jacobi preconditioner due to its simplicity and low computational overhead. It is defined as the diagonal of the system matrix ($M = \text{diag}(A)$) and its inverse $M^{-1}$ is obtained by taking the reciprocals of the diagonal elements. Since it is not available in vendor libraries, we implemented it ourselves.

\subsubsection{Communication}
\label{section:communication}
Parallel dot product operations require collective allreduce operations
to sum up the partial dot product results from every GPU into a single global value
to ensure the mathematical correctness of the distributed algorithm.

The second type of the communication required in pipelined CG and pipelined BICGSTAB concerns SpMV. 
More specifically, in a distributed SpMV ($y = Ax$)
with row-wise partitioning, each GPU computes a partial output vector from
its assigned rows. Because GPUs only require elements of the vector $x$ that 
correspond to the non-zero columns in their assigned matrix partition, 
SpMV generates an irregular point-to-point communication pattern. 
To manage this efficiently and hide network latency, the assigned matrix 
partition is conceptually divided into local and remote components, often referred to as \textit{diagonal} and \textit{off-diagonal} parts, respectively. 
Each GPU first multiplies the local (diagonal) matrix component with locally available $x$ values. 
Simultaneously, it initiates P2P communication to fetch the missing remote $x$ values. 
Once the transfer is completed, the GPU multiplies the off-diagonal matrix 
component with the newly received data. 
This strategy allows us to overlap 
local computation with halo communication, and we utilize it throughout our implementations. 






\subsection{Two-Stream Approach}
\begin{figure*}[t]
    \centering
    \includegraphics[width=0.94\linewidth, keepaspectratio]{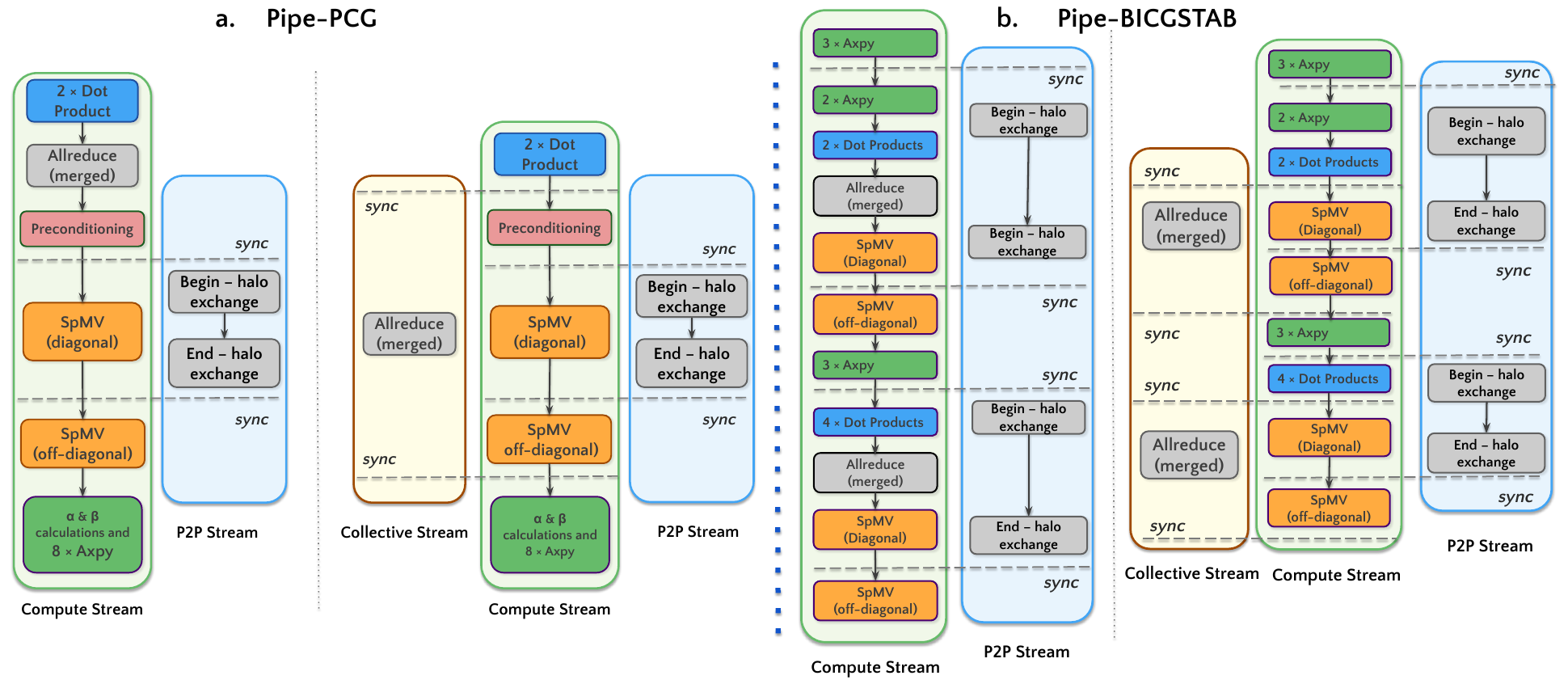}
    \caption{Two-stream vs Three-stream design of Pipelined Preconditioned CG and Pipelined BICGSTAB}
    \label{fig:2-3stream}
\end{figure*}

We exploit the opportunity to overlap the irregular P2P communication of distributed memory SpMV with computation on the local part of the sparse matrix
using GPU streams, as illustrated in Figure~\ref{fig:2-3stream}. Our implementation builds on a state-of-the-art approach~\cite{trotter2025cpu}, which combines CUDA streams with NCCL to avoid host-side synchronizations.
Since NCCL is stream-aware, GPU kernels for packing, sending, receiving and unpacking messages can be queued asynchronously from the CPU without incurring CPU-GPU synchronization overhead. Dependencies between the {\itshape P2P} stream and {\itshape compute} stream are handled asynchronously through CUDA events (\texttt{cudaEventRecord} and \texttt{cudaStreamWaitEvent}), which further eliminates CPU-GPU synchronization and hides kernel launch latencies.

As shown in Figure~\ref{fig:2-3stream}-a for Pipe-PCG, once the preconditioning step completes, synchronization is established by asynchronously recording an event on the compute stream and making the P2P stream wait on that event. Conversely, the P2P stream records an event upon completing the P2P communication, and the compute stream thus waits on this event before proceeding with the off-diagonal part of the SpMV.
Similarly in Pipe-BICGSTAB as shown in Figure~\ref{fig:2-3stream}-b, once the first group of fused AXPY operations complete, compute stream signals to P2P stream and initiates the halo exchange operations.
This ensures that the required vector data is ready for both the SpMV operation and the P2P transfer. Non-blocking point-to-point communication is implemented using grouped calls to \texttt{ncclSend} and \texttt{ncclRecv}, enclosed by \texttt{ncclGroupStart} and \texttt{ncclGroupEnd}. 

\subsection{Three-Stream Approach}

In the two-stream strategy, the collective communication required for the dot products remains a synchronization bottleneck, unable to be overlapped with other data transfers or computations. 
To achieve the triple overlap, we propose a novel three-stream implementation of Pipe-PCG and Pipe-BICGSTAB. As illustrated in Fig.~\ref{fig:2-3stream}, the collective operations  are offloaded to a dedicated \textit{Collective} stream. The core mathematical computations  are maintained in the primary \textit{Compute} stream, and the P2P data transfers remain isolated in their own dedicated \textit{P2P} stream. Listing \ref{lst:cg-multi-gpu} shows the pseudo-code of the multi-GPU three-stream implementations for Pipe-PCG and Pipe-BICGSTAB.

\begin{figure}
    \centering
    \includegraphics[width=\linewidth, height=0.5\textheight, keepaspectratio]{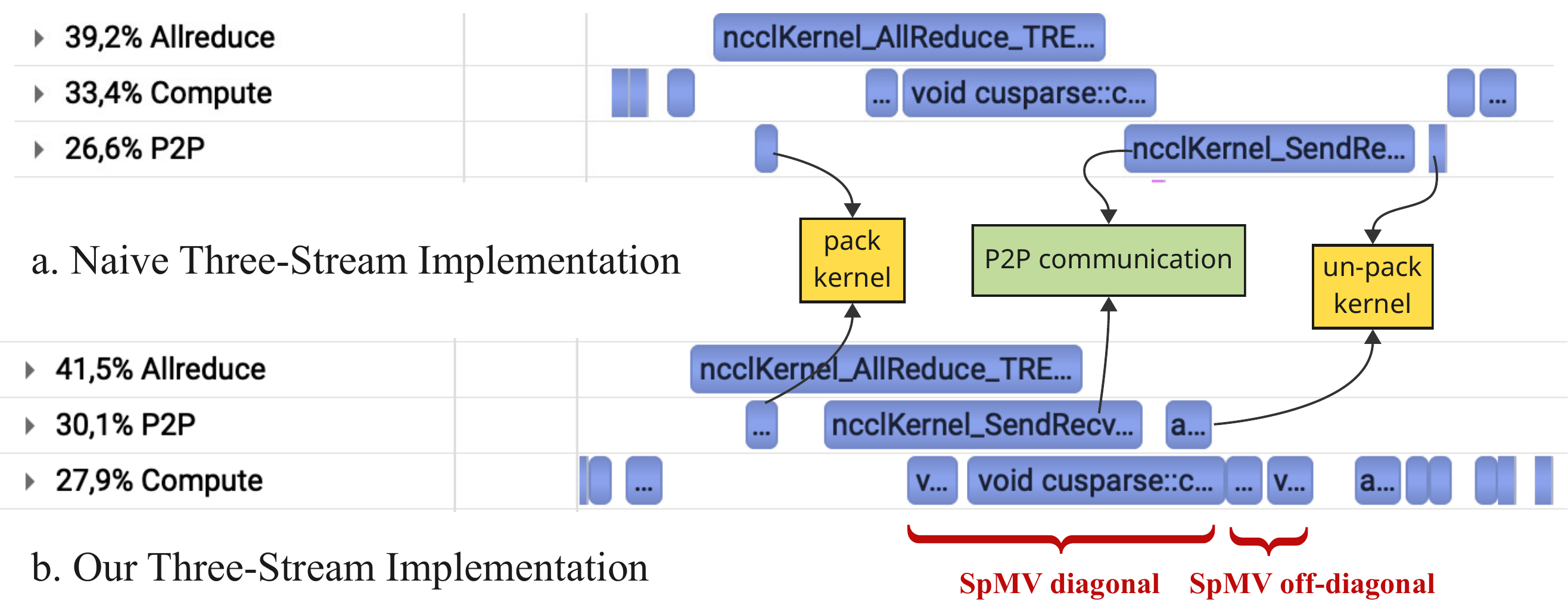}
    \caption{Nsight Systems profiling results of naive three-stream  vs our three-stream implementation}
    \label{fig:3-stream-profiling}
\end{figure}

However, placing the collective in a separate  stream is not sufficient to achieve true overlap with P2P transfers. Fig.~\ref{fig:3-stream-profiling}-a shows the profiling results for a naive three-stream design based on a single NCCL communicator. In this case, only one communication operation overlaps with computation, while the others remain queued until the active transfer finishes. As discussed in Sec.~\ref{section:communication}, irregular P2P communication is required to fetch remote $x$ values, which are packed before transfer and unpacked on arrival. In the naive design, however, NCCL serializes the two communication operations and, in this case, prioritizes the collective. Consequently, in Pipe-PCG implementation, the P2P transfer is delayed until the Allreduce completes, which in turn causes the off-diagonal SpMV to execute much later than the diagonal SpMV because the required remote $x$ values are not yet available.
Similarly in Pipe-BICGSTAB, as shown in Fig.~\ref{fig:2-3stream}-b, we start overlapping halo exchange operations after the first group of AXPYs, and since NCCL serializes communications in the background, Allreduce delays until the completion of halo exchange operations and we waste the opportunity of having a wider overlap window. Rather than overlapping diagonal SpMV, Allreduce and P2P  together we end up overlapping halo exchanges with AXPYs, DOT products and diagonal SpMV; and Allreduce with off-diagonal SpMV. 



To address this problem, separate communicators must be used for different communication operations. There are two ways to achieve this. The first is to create two independent  communicators and use each for a different type of communication. However, this approach requires additional resource management as well as careful handling of possible deadlocks due to the 
order of host-side launches. The second is to split an existing communicator using NCCL’s {\ttfamily ncclCommSplit} feature. {\ttfamily ncclCommSplit} partitions an existing communicator into multiple sub-communicators, enabling different communication operations to proceed on separate communicators and thereby allowing true overlap. This overlap is illustrated in Figure~\ref{fig:3-stream-profiling}-b in the  profiling results. In our implementation, we adopt {\ttfamily ncclCommSplit} because it simplifies resource management internally and avoids possible deadlocks. 

%% file: sections/codes/three-stream-code.tex
\lstdefinestyle{pseudocode}{%
  mathescape=true,
  keepspaces=true,
  columns=fullflexible,
  numbers=left,
  numberstyle=\footnotesize,
  basicstyle=\ttfamily\footnotesize,
  keywordstyle=\color{black}\bfseries,
  keywords={,Input, Output, Return, Exit, In, If, Then, Else, End, For, Do, }
  xleftmargin=.01\textwidth,
  moredelim=**[is][\color{OliveGreen}]{@}{@},
  moredelim=**[is][\color{NavyBlue}]{/}{/},
  moredelim=**[is][\color{Plum}]{!}{!},
  moredelim=**[is][\color{Brown}]{~}{~}}

\begin{lstlisting}[%
  frame=single,%
  style=pseudocode,%
  caption={Pseudo-code of the multi-GPU three-stream Pipe-PCG implementation. Colors indicate the stream on which each operation is executed: the \textcolor{OliveGreen}{\itshape compute stream} in green, the \textcolor{Plum}{\itshape P2P stream} in purple, the \textcolor{Brown}{\itshape collective stream} in brown, and \textcolor{NavyBlue}{\itshape host-side CPU operations} in blue.},
  label={lst:cg-multi-gpu},%
  float]
Input: $\boldsymbol{A}^{(p)}=\boldsymbol{\bar{A}}^{(p)}+\boldsymbol{\hat{A}}^{(p)}$, $\boldsymbol{b}^{(p)}$, $\boldsymbol{x}_0^{(p)}$, $\boldsymbol{M}$, $\epsilon$
$\boldsymbol{r}_0 \gets \boldsymbol{b} - \boldsymbol{A}\boldsymbol{x}_0$, $\boldsymbol{u}_0 \gets \boldsymbol{M}^{-1}\boldsymbol{r}_0$, $\boldsymbol{w}_0 \gets \boldsymbol{A}\boldsymbol{u}_0$
For $i=0,1,2,\dots$ Do
  @GPU DOT@:   $\gamma_i^{(p)} \gets \boldsymbol{r_i}\cdot{}\boldsymbol{u_i}$
  @GPU DOT@:   $\delta^{(p)}$ $\gets$ $\boldsymbol{w}_i\cdot{}\boldsymbol{u}_i$
  ~SYNC~:      Wait for dot products to complete
  ~ALLREDUCE~: $\gamma_i \gets \sum_p \gamma_i^{(p)}$; $\delta \gets \sum_p \delta^{(p)}$ (non-blocking)
  @GPU Preconditioner@:   $\boldsymbol{m}_i$ $\gets$ $\boldsymbol{M}^{-1}\boldsymbol{w}_i$
  !SYNC!:      Wait until $\boldsymbol{m}_i$ is ready
  !P2P PACK!:  Pack halo $\boldsymbol{\hat{m}_i}$ for sending
  !P2P COMM!:  Start sending halo $\boldsymbol{\hat{m}}_i$ to remote GPUs
  @GPU SpMV@:  $\boldsymbol{n}_i \gets \boldsymbol{\bar{A}}^{(p)}*\boldsymbol{m}_i$ (Diagonal)
  @SYNC@:      Wait until $\boldsymbol{\hat{m}}_i$ is received
  @GPU SpMV@:  $\boldsymbol{n}_i \gets \boldsymbol{n}_i+\boldsymbol{\hat{A}}^{(p)}*\boldsymbol{\hat{m}}_i$ (Off-diagonal)
  @SYNC@:      Wait until allreduce is done
  /CPU/:       If $||\boldsymbol{r}_i|| \leq \epsilon||\boldsymbol{r}_0||$ Then Exit: Converged
  @GPU FUSED AXPY@:
        If $i == 0$: $\alpha_0 \gets \gamma_0 / \delta$, $\beta_0 = 0$
        Else: $\alpha_i \gets \gamma_i / (\delta - \beta_i \gamma_i / \alpha_{i-1})$, $\beta_i \gets \gamma_i /\gamma_{i-1}$
        $\boldsymbol{z}_i \gets \boldsymbol{n}_i + \beta_i \boldsymbol{z}_{i-1}$
        $\boldsymbol{q}_i \gets \boldsymbol{m}_i + \beta_i \boldsymbol{q}_{i-1}$
        $\boldsymbol{s}_i \gets \boldsymbol{w}_i + \beta_i \boldsymbol{s}_{i-1}$
        $\boldsymbol{p}_i \gets \boldsymbol{u}_i + \beta_i \boldsymbol{p}_{i-1}$
        $\boldsymbol{x}_{i+1} \gets \boldsymbol{x}_i + \alpha_i \boldsymbol{p}_i$
        $\boldsymbol{r}_{i+1} \gets \boldsymbol{r}_i - \alpha_i \boldsymbol{s}_i$
        $\boldsymbol{u}_{i+1} \gets \boldsymbol{u}_i - \alpha_i \boldsymbol{q}_i$
        $\boldsymbol{w}_{i+1} \gets \boldsymbol{w}_i - \alpha_i \boldsymbol{z}_i$
End For
\end{lstlisting}

%% file: sections/experiments.tex
\section{Experiments and Evaluation}
\label{sec:experiments}
We compare our proposed three-stream implementation with the state-of-the-art two-stream Pipe-PCG and Pipe-BiCGStab  and the PETSc's implementations, which are based on NCCL and CUDA-aware MPI, respectively.
\subsection{Experimental Setup}

\begin{table}[t]
\centering
\caption{System configurations.}
\label{tab:system-configuration}
\scriptsize
\setlength{\tabcolsep}{3pt}
\renewcommand{\arraystretch}{1.15}
\begin{tabularx}{\columnwidth}{@{}l >{\raggedright\arraybackslash}X >{\raggedright\arraybackslash}X@{}}
\toprule
 & \textbf{MareNostrum5 (ACC)} & \textbf{Leonardo (Booster)} \\
\midrule
CPU        & 2$\times$ Xeon Platinum 8460Y+ (40c) & 1$\times$ Xeon Platinum 8358 (32c) \\
GPU        & 4$\times$ H100, 64\,GB HBM2e & 4$\times$ A100 SXM4, 64\,GB @ 1{,}640\,GB/s \\
Intra-node & NVLink 4.0, 300\,GB/s bi-dir.\ per pair & NVLink 3.0, 200\,GB/s per pair \\
NIC        & ConnectX-7, 200\,Gb/s uni-dir.\ per GPU & 2$\times$ dual-port HDR100 ConnectX-6, 400\,Gb/s \\
Network    & NDR IB, 3-layer fat-tree~\cite{banchelli2025introducing} & IB~\cite{turisini2023leonardo} \\
Software   & HPC SDK 25.1, CUDA 12.6, HPC-X 2.21, NCCL 2.18.5 & HPC SDK 24.3, CUDA 12.3, HPC-X 2.17.1, NCCL 2.18.5 \\
\bottomrule
\end{tabularx}
\end{table}

\paragraph{\textbf{Supercomputers}}
We evaluated our implementations on two leading  supercomputers, MareNostrum5 and Leonardo. Table~\ref{tab:system-configuration} details their configurations and software stacks.


\paragraph{\textbf{Matrices}}

We use two groups of test instances. The first is taken from Suitesparse \cite{davis2011university} where we select the six largest positive definite symmetric real-valued matrices. 
Since they are not large enough to scale beyond 64 GPUs, we also add three very large test instances. 
The first is a matrix from discretising a 3D Poisson equation on a unit cube geometry. 
The other two are derived from a 3D cardiac monodomain simulation of the human heart 
at different resolutions \cite{martinez-navarro2019}. They represent an unstructured, 
tetrahedral mesh based and are thus highly irregular and very sparse. 
Table \ref{tab:matrices} lists the instances, their sizes, and their nonzero 
distribution by row. 

 
\input{tables/matrices}
\input{tables/speed-ups}

\begin{figure*}
    \centering
    \includegraphics[width=\linewidth, height=0.5\textheight, keepaspectratio]{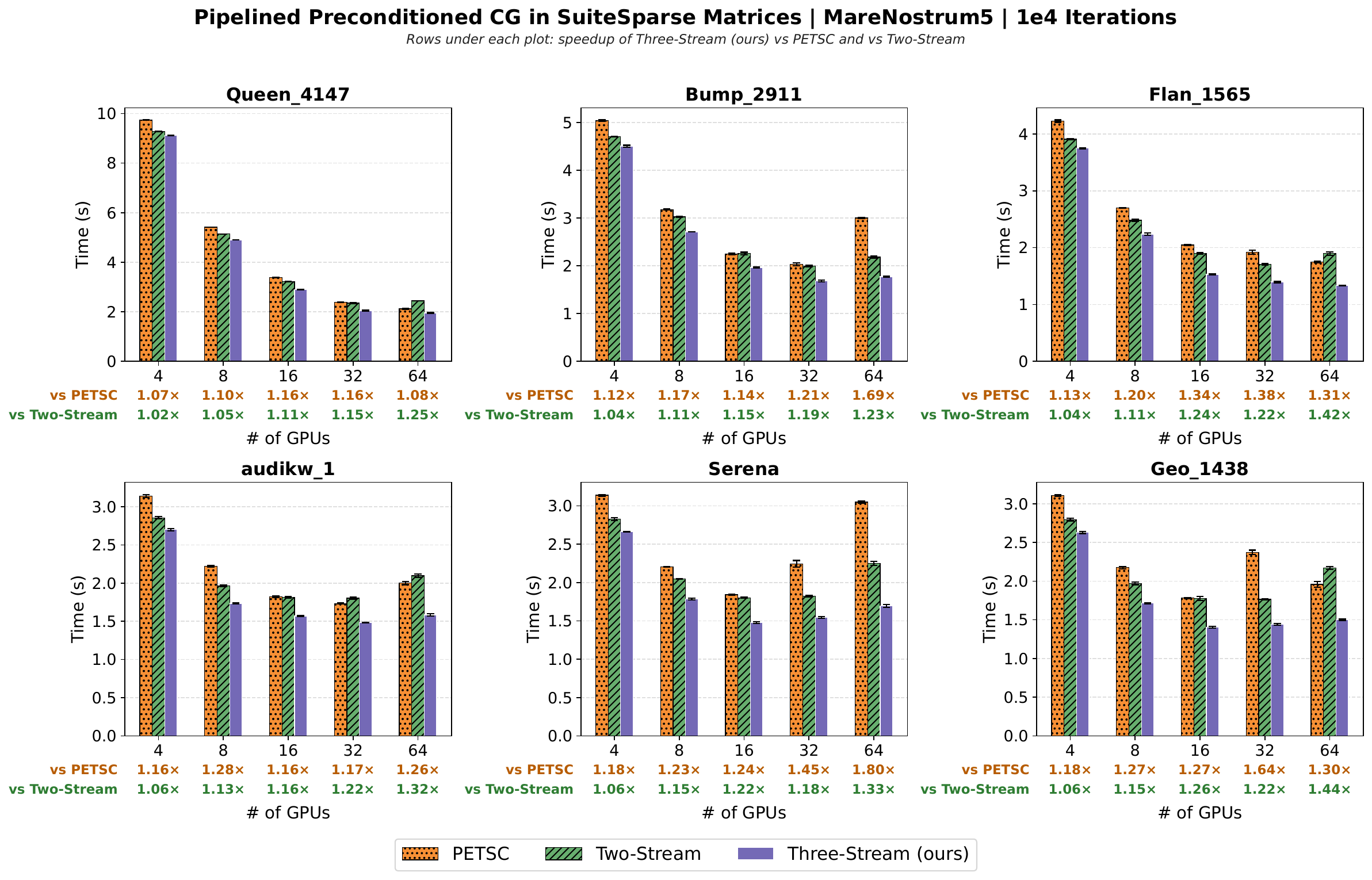}
    \caption{Performance of the proposed Three-Stream design compared with PETSc and the state-of-the-art Two-Stream implementation \cite{trotter2025cpu} on SuiteSparse matrices on the \textbf{MareNostrum5} supercomputer. Numbers above the Three-Stream bars show the speedup over PETSc (orange) and Two-Stream implementation (green).}
    \label{fig:MN5_suitesparse}
\end{figure*}

\begin{figure*}
    \centering
    \includegraphics[width=\linewidth, height=0.5\textheight, keepaspectratio]{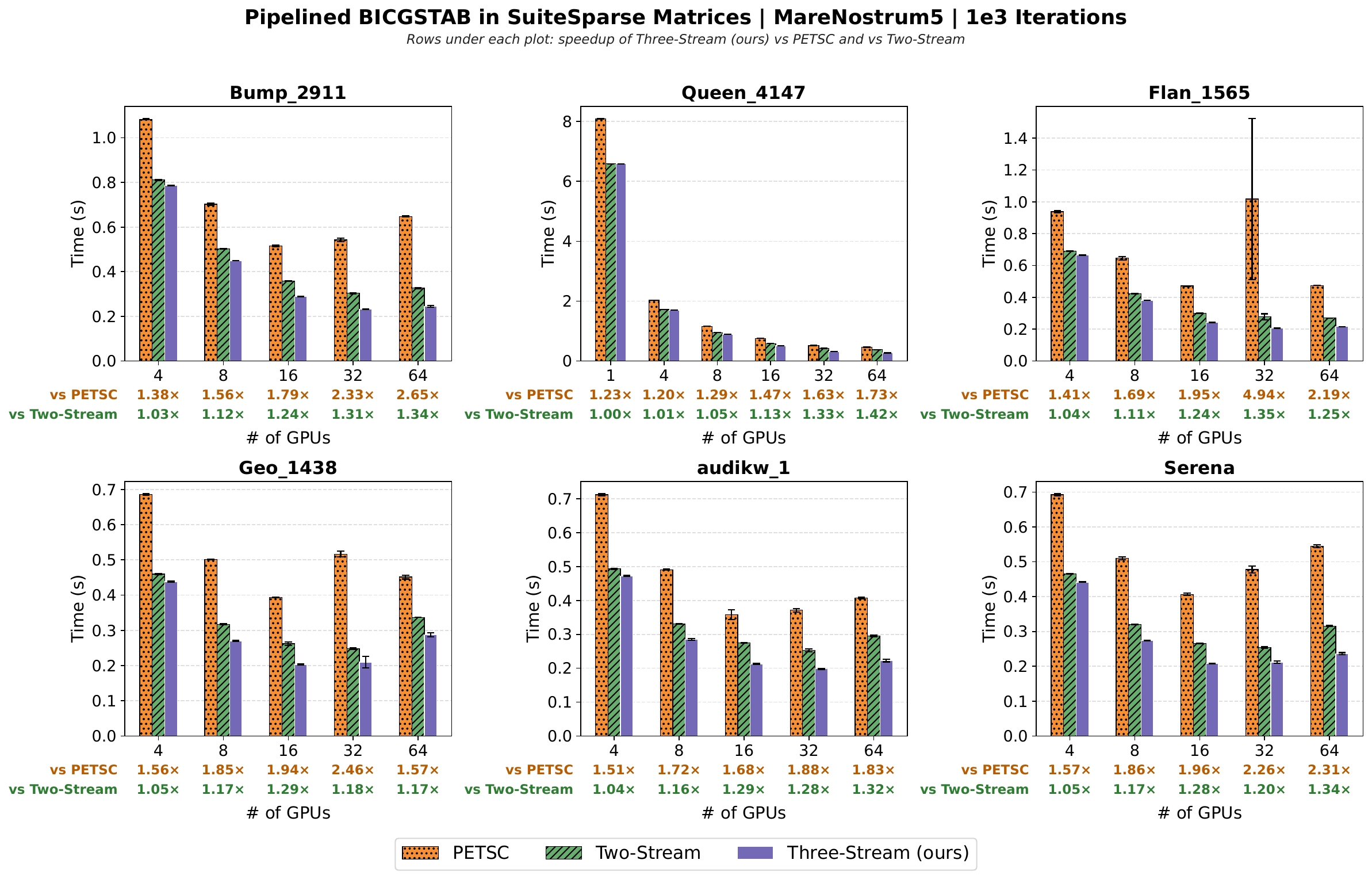}
    \caption{Performance of the proposed Three-Stream design for Pipe-BICGSTAB compared with PETSc and the state-of-the-art Two-Stream implementation \cite{trotter2025cpu} on SuiteSparse matrices on the \textbf{MareNostrum5} supercomputer. Numbers above the Three-Stream bars show the speedup over PETSc (orange) and Two-Stream implementation (green). 
    }
    \label{fig:MN5_suitesparse_bicgstab}
\end{figure*}

\paragraph{\textbf{Measurement}}

Execution time is measured on the host CPU for each MPI process using  {\ttfamily clock\_gettime}. To ensure accurate timings, all solvers execute 10 warmup iterations, followed by an MPI barrier and CPU-GPU synchronization (e.g., {\ttfamily cudaDeviceSynchronize}), before the timer begins. A final CPU-GPU synchronization occurs before stopping the timer to guarantee that all GPU tasks are fully completed. We then report the maximum execution time across all participating  processes.

We evaluate all solvers including PETSc within the same job submission for a given matrix and GPU count. This ensures the use of identical matrix partitioning, compute nodes, and GPUs. We repeat each measurement three times and report the average with error bars showing the range across three runs. To avoid outliers across the iterations in the solver loop, we ran each solver for 10,000 iterations for Pipe-PCG runs and 1,000 iterations for Pipe-BiCGStab.
Finally, for an in-depth analysis of the two-stream and three-stream implementations, we conducted additional fine-grained measurements of individual GPU kernel times using  {\ttfamily cudaEventRecord}.

\subsection{Experimental Results}

We first present experimental results for the Suitesparse matrices. 
Figure \ref{fig:MN5_suitesparse} shows results on MareNostrum5. 
In all cases, Three-Stream is ahead of the Two-Stream implementation. Furthermore, the gap widens consistently with growing number of GPUs, from 2--6\% at 4 GPUs to 18--47\% at 64 GPUs, with only Geo\_1438 on MareNostrum5 constituting an outlier due to unstable results (as evidenced by the large variability between the runs). 


Since the cost of \textit{allreduce} grows with the number of GPUs, so does the benefit of overlapping it with P2P communication, confirming that the Three-Stream design works as intended. 
While the relative benefit should be larger for shorter total running times, the 64-GPU running times are too similar across instances for this effect to be visible. 

In addition to the figures, Table \ref{tab:speedups} shows the speedups including the speedup versus PETSc, which uses GPU-aware MPI rather than NCCL. While PETSc is almost always slower than the stream-based implementations, the difference is much more pronounced on Leonardo, where Three-Stream is up to 93\% faster
with a geometric mean of $1.72$ whereas the speedup over Two-Stream is much lower. 

All matrices show some performance scaling initially, but only for the largest instance, Queen\_4147, the scaling is close to linear. None of the instances show any noticeable improvement at 64 GPUs, which implies that our experiments have reached the limits of scalability for these matrices.

For Pipe-BiCGStab, Three-Stream again outperforms Two-Stream on every instance and GPU count (Figure~\ref{fig:MN5_suitesparse_bicgstab}), with the margin growing from 1--5\% at 4 GPUs to 17--42\% at 64 GPUs, where the geometric-mean speedup reaches $1.30\times$ over Two-Stream and $2.01\times$ over PETSc.



\begin{figure*}
    \centering
    \includegraphics[width=\linewidth, height=0.5\textheight, keepaspectratio]{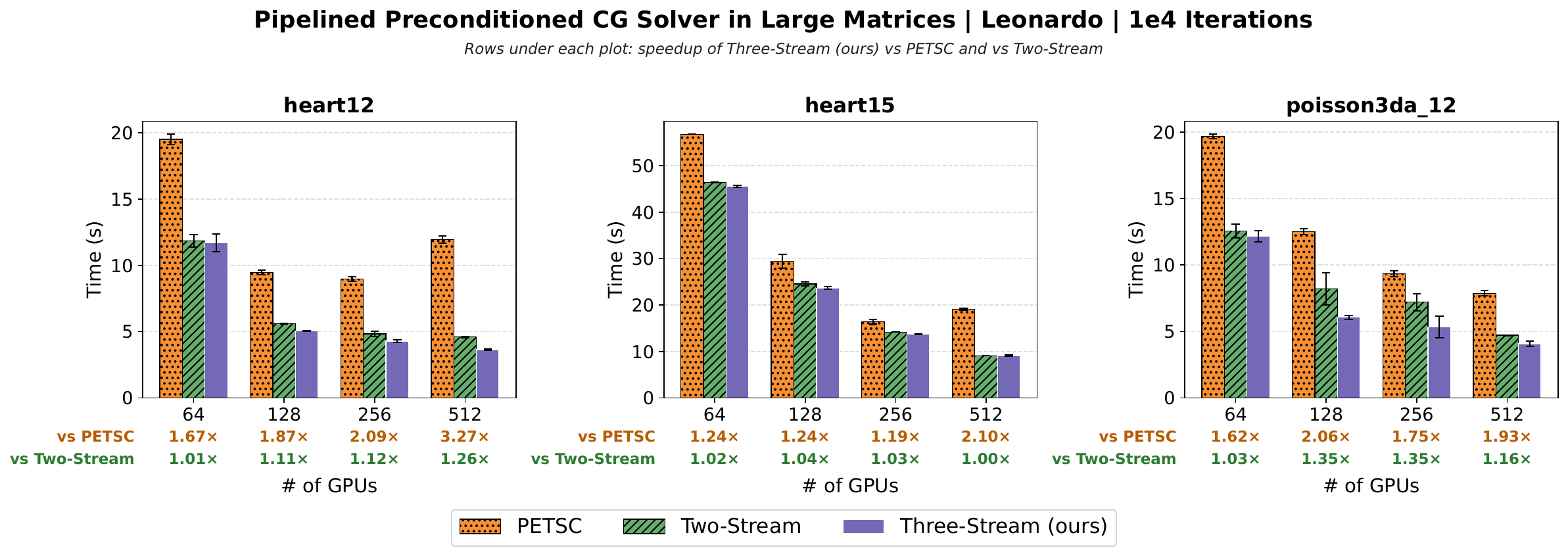}
    \caption{Performance of the proposed Three-Stream design compared with PETSc and the state-of-the-art Two-Stream implementation \cite{trotter2025cpu} on larger matrices on the \textbf{Leonardo} supercomputer. Numbers above the Three-Stream bars show the speedup over PETSc (orange) and Two-Stream implementation (green).}
    \label{fig:Leonardo_large_matrices}
\end{figure*}
\begin{figure*}
    \centering
    \includegraphics[width=\linewidth, height=0.5\textheight, keepaspectratio]{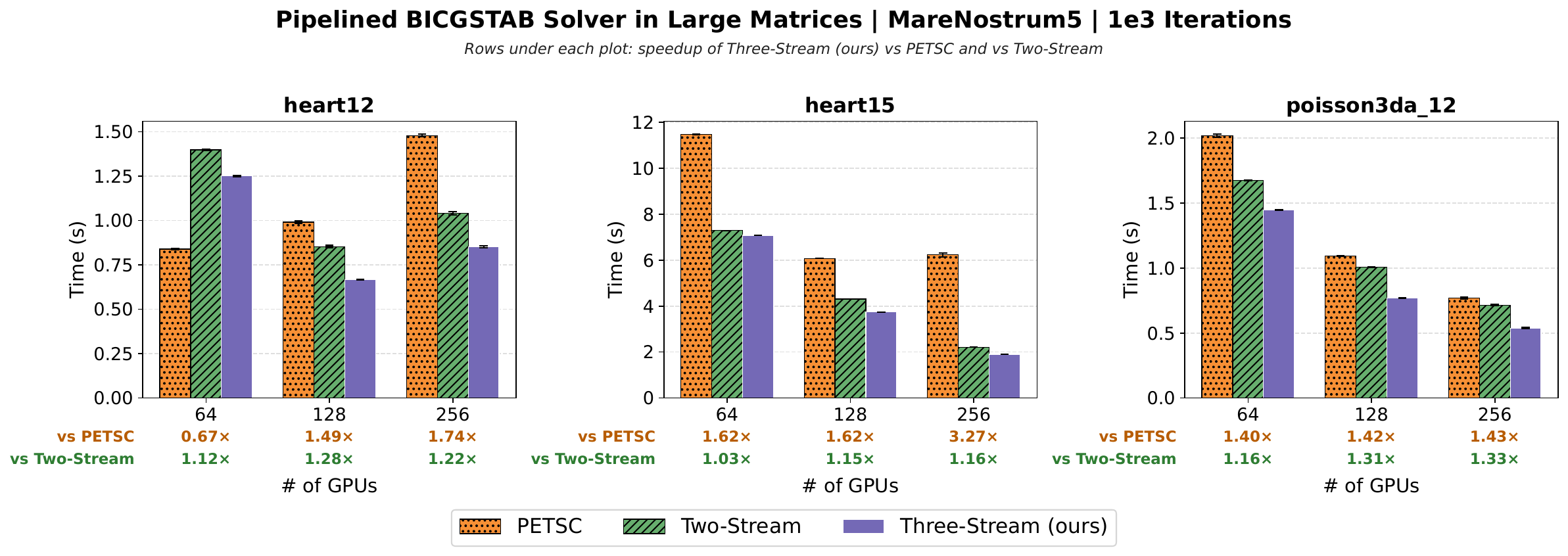}
    \caption{Performance of the proposed Three-Stream design for Pipe-BiCGStab compared with PETSc and the state-of-the-art Two-Stream implementation \cite{trotter2025cpu} on larger matrices on the \textbf{MareNostrum5} supercomputer. Numbers above the Three-Stream bars show the speedup over PETSc (orange) and Two-Stream implementation (green).}
    \label{fig:MN5_large_matrices_bicgstab}
\end{figure*}

For the large matrix instances, Figure~\ref{fig:Leonardo_large_matrices} shows Pipe-PCG results on Leonardo, while Figure~\ref{fig:MN5_large_matrices_bicgstab} shows the Pipe-BICGStab results on MareNostrum5. Since these problems require more computation, the relative communication overhead is small at lower GPU counts, and the three-stream advantage starts modestly, but grows o $1.52\times$, $1.06\times$, and $1.50\times$ at 256 GPUs on heart12, heart15, and poisson3da\_12, respectively as the collective communications become more expensive relative to the computation, leaving more latency for the third stream to hide.

Finally, PETSc is highly competitive on these large instances, even beating our design on heart12 at 64 GPUs (where we reach only $0.67\times$ of its performance). This exposes the pipelined approach's weak point: massive local work, cheap collectives, and unamortized extra vector operations. At 128 and 256 GPUs, however, this effect disappears, and our three-stream design outpaces PETSc by $1.49\times$ and $1.74\times$ on the same matrix.

\subsection{Performance Analysis}

To better understand the scaling behavior observed in the last section, we profiled experiments for Pipe-PCG conducted on the MareNostrum5 supercomputer using 4 and 64 GPUs for the Queen\_4147 matrix and Flan\_1565 matrices. We measure the time spent in compute kernels directly using CUDA events. The same method cannot be used to provide isolated measurements of the communication time, because such measurements would also include effects of synchronization and load imbalance among the communicating GPUs. Instead, we estimate the overall time spent in communication (i.e., {\itshape non-compute}) by subtracting the times for the compute kernels from the total solver time.

The effective memory bandwidth of the local SpMV drops from 82\% of the theoretical peak at 4 GPUs to 65\% at 64 GPUs for Queen\_4147, and from 56\% to 9.8\% for Flan\_1565, indicating that the latter matrix is too small to run efficiently at this scale. For reference, our BabelStream measurements show that a single GPU reaches about 90\% of the theoretical maximum: SpMV approaches this value only under ideal conditions, and not all matrices can do so without reordering~\cite{trotter2023bringing}.


The remaining compute kernels (preconditioner, diagonal SpMV, off-diagonal SpMV, and AXPYs) hardly scale with the number of GPUs. For the off-diagonal SpMV, this is because the number of halo elements to be communicated grows with the number of GPUs. For the others, the work per kernel invocation becomes so small at 64 GPUs that launch overheads dominate. Since these overheads, like Allreduce, are independent of the instance, the 64-GPU running times cluster around 1.5 secs for all instances except the much larger Queen\_4147 ($\sim$2 secs).

For communication, even at 4 GPUs on Queen\_4147, Three-Stream communication is 28\% faster than Two-Stream. However, since communication accounts for only 14\% (Two-Stream) and 12\% (Three-Stream) of the total time, the overall improvement is marginal. At 64 GPUs, in contrast, communication is \%48 faster and accounts for 55\% (Two-Stream) and 46\% (Three-Stream) of the total time, yielding the overall speedup of 24\%.

%% file: tables/matrices.tex
\begin{table}
  \centering
  \caption{Sparse matrices used in experiments. }
\scriptsize
  \begin{tabular}{lrrr}
    \toprule
    Matrix            & Rows          & Nonzeros      & Nonzeros/Row \\
    \midrule
    Queen\_4147       &     4,147,110 &   316,548,962 &  76.33 \\
    Bump\_2911        &     2,911,419 &   127,729,899 &  43.87  \\
    Flan\_1565        &     1,564,794 &   114,165,372 &  72.96  \\
    audikw\_1         &       943,695 &    77,651,847 &  82.28  \\ 
    Serena            &     1,391,349 &    64,131,971 &  46.09  \\
    Geo\_1438         &     1,437,960 &    60,236,322 &  41.89  \\

    \midrule
    poisson3da\_12      & 144,515,663 & 2,239,624,249 & 15.5 \\
    heart15           &   874,424,445 & 7,185,411,541 & 8.2 \\
    heart12           &   139,589,939 & 2,144,835,461 & 15.4 \\
    \bottomrule
  \end{tabular}
  \label{tab:matrices}
\end{table}

%% file: tables/speed-ups.tex
\begin{table}[t]
\centering
\caption{Speed-ups of three-stream (ours) over PETSc and two-stream for Pipe-PCG on MareNostrum5 (H100) and Leonardo (A100) using 64 GPUs.}
\label{tab:speedups}
\resizebox{\columnwidth}{!}{
\begin{tabular}{l c c c c}
\toprule
 & \multicolumn{2}{c}{\textbf{MareNostrum5 (H100)}} & \multicolumn{2}{c}{\textbf{Leonardo (A100)}} \\
\cmidrule(lr){2-3} \cmidrule(lr){4-5}
\textbf{Matrix} &  \textbf{vs.\ PETSc} & \textbf{vs.\ 2-stream} &  \textbf{vs.\ PETSc} & \textbf{vs.\ 2-stream} \\
\midrule
Queen\_4147    & 1.08$\times$ & 1.25$\times$   & 1.93$\times$ & 1.12$\times$ \\
Bump\_2911      & 1.69$\times$ & 1.23$\times$   & 1.69$\times$ & 1.18$\times$ \\
Flan\_1565      & 1.31$\times$ & 1.42$\times$   & 1.56$\times$ & 1.47$\times$ \\
audikw\_1      & 1.26$\times$ & 1.32$\times$  & 1.82$\times$ & 1.31$\times$ \\
Serena          & 1.80$\times$ & 1.33$\times$   & 1.66$\times$ & 1.26$\times$ \\
Geo\_1328       & 1.30$\times$ & 1.44$\times$   & 1.68$\times$ & 1.09$\times$ \\
\midrule
Geomean         & 1.38	&   1.33  &	1.72 &	1.23\\
\bottomrule
\end{tabular}}
\label{tab:speedups}
\end{table}

%% file: sections/related_works.tex
\section{Related Works}
\label{sec:related-works}

The Pipelined-CG was introduced to reduce the cost of global synchronization in Krylov methods. Ghysels and Vanroose reformulated PCG so that a single nonblocking reduction is issued per iteration and overlapped with the matrix-vector product and preconditioner \cite{ghysels2014hiding}. Later work extended this idea with deep pipelines that hide a reduction behind multiple future SpMVs when one iteration does not provide sufficient  work \cite{deeppipelineCG}. More recent pipelined-PCG variants further reduce reduction frequency by overlapping one allreduce with work from two iterations \cite{tiwari-DistributedMemCG}. However, this line of work only targets global dot product latency. Moreover, although non-blocking collectives were introduced in  MPI-3, their effectiveness on CPU clusters depends on factors such as asynchronous progress and runtime support \cite{Denis2022}.

A separate line of work focuses on halo/P2P communication overlap in iterative solvers and multigrid methods. These approaches aim to overlap halo exchange in the SpMV or smoother/preconditioner phases with computation. For example, Nakajima et al.~combine halo overlap with OpenMP dynamic loop scheduling \cite{Nakajima2022OpenMP}, and later study process/thread allocation in hybrid MPI+OpenMP settings to optimize communication-computation overlap in a CG solver with a multigrid preconditioner \cite{Nakajima2025CCOverlapMG}. Thus, these works do not address the joint scheduling of halo exchange and collective reductions.

On GPU systems,  Bernaschi et al.~combine communication-reduced CG with overlap in the multi-GPU SpMV \cite{PasquaCG}. Ismayilov et al.~use persistent kernels, device-side barriers, and device-initiated communication to remove the CPU from the critical path in multi-GPU CG \cite{cpu-free}. Trotter et al.~\cite{trotter2025cpu} compare CPU- and GPU-initiated communication strategies for CG and pipelined CG across GPU-aware MPI, NCCL/RCCL, and NVSHMEM. However none of these works consider triple overlap in their implementations.

The closest prior work to ours is Selvitopi et al.~\cite{selvitopi}, who propose a rearrangement for iterative solvers in which irregular SpMV P2P messages are embedded into a collective communication phase so that P2P and collective communication can be performed in a single phase. Although this mixes the two communication classes, it is not a pipelined-CG formulation that treats collective reduction, halo exchange, and SpMV as three concurrent components.

Outside CG, MPI+ULT \cite{MPI-ULT} argues that user-level threads can overlap computation with both P2P and collective MPI communication, but it does not demonstrate a true three-way overlap of collective, P2P, and compute. Lina \cite{Lina} studies distributed Mixture-of-Experts training, where all-to-all and allreduce coexist with computation on separate streams. It shows that naive concurrency between these communication classes can hurt performance due to network contention, and therefore prioritizes all-to-all over allreduce. Our approach differs in that we overlap collective communication with P2P exchange, yielding a different communication pattern and less pressure on the network bandwidth.

%% file: sections/AD.tex
\newcommand{\cmd}[1]{{\small\ttfamily #1}}

\twocolumn[%
{\begin{center}
\Huge
Appendix: Artifact Description/Artifact Evaluation
\end{center}}
]


\appendixAD

\section{Overview of Contributions and Artifacts}

\subsection{Computational Artifacts}

All computational artifacts have been archived on Zenodo:

\begin{description}
\item[$A_1$] https://doi.org/10.5281/zenodo.22973787
\item[$A_2$] https://doi.org/10.5281/zenodo.22973707
\end{description}

The main computational artifacts are included in $A_1$. Additional input files are provided in $A_2$, related to the matrix poisson3da\_12, which is used for some of the large-scale experiments in the paper.

\section{Artifact Identification}

\newartifact

\artrel

Artifact $A_1$ includes:
\begin{enumerate}
    \item Source code for the solvers, which implement the various two-stream and three-stream multi-GPU CG solvers that are used for the performance benchmarks presented in the paper.

    \item This dataset contains the row partitioning files for six matrices selected from the SuiteSparse Collection (Davis and Hu, 2011): Bump\_2911, Geo\_1438, Flan\_1565, Queen\_4147, Serena, and audikw\_1. Using the METIS graph partitioner (Karypis and Kumar, 1998), each matrix was partitioned into 4, 8, 16, 32 and, 64 parts.
    
    \item Included here are the job submission scripts used to benchmark the multi-GPU CG solvers (two- and three-stream implementattions and PETSc) on the MareNostrum 5, and Leonardo clusters.
\end{enumerate}
\artexp
By executing the two-stream, three-stream, and PETSc experiments using the provided source code and open datasets, users can reproduce the solver runtimes and speedups illustrated in the paper.

\arttime
The anticipated computational time for the SuiteSparse matrix experiments ranges from 30 seconds to 15 minutes for each combination of solver, matrix, and GPU count. For the large-scale experiments, executing each solver and GPU configuration requires approximately 10 to 15 minutes for the "poisson3da\_12", "heart12" and "heart15" matrices on the MareNostrum 5 and Leonardo clusters. 

\artin

Hardware and their configurations used in the experiments can be seen in Table~\ref{tab:system-configuration}.
\begin{table}[t]
\centering
\caption{System configurations.}
\label{tab:system-configuration}
\scriptsize
\setlength{\tabcolsep}{3pt}
\renewcommand{\arraystretch}{1.15}
\begin{tabularx}{\columnwidth}{@{}l >{\raggedright\arraybackslash}X >{\raggedright\arraybackslash}X@{}}
\toprule
 & \textbf{MareNostrum5 (ACC)} & \textbf{Leonardo (Booster)} \\
\midrule
CPU        & 2$\times$ Xeon Platinum 8460Y+ (40c) & 1$\times$ Xeon Platinum 8358 (32c) \\
GPU        & 4$\times$ H100, 64\,GB HBM2e & 4$\times$ A100 SXM4, 64\,GB @ 1{,}640\,GB/s \\
Intra-node & NVLink 4.0, 300\,GB/s bi-dir.\ per pair & NVLink 3.0, 200\,GB/s per pair \\
NIC        & ConnectX-7, 200\,Gb/s uni-dir.\ per GPU & 2$\times$ dual-port HDR100 ConnectX-6, 400\,Gb/s \\
Network    & NDR IB, 3-layer fat-tree~\cite{banchelli2025introducing} & IB~\cite{turisini2023leonardo} \\
Software   & HPC SDK 25.1, CUDA 12.6, HPC-X 2.21, NCCL 2.18.5 & HPC SDK 24.3, CUDA 12.3, HPC-X 2.17.1, NCCL 2.18.5 \\
\bottomrule
\end{tabularx}
\end{table}
\artinpart{Software}

Introduce all required software packages, including the computational artifact. For each software package, specify the version and provide the URL.

The following software is used on all systems:
\begin{itemize}
\item CG Solvers:\newline (https://doi.org/10.5281/zenodo.22973787)
\item PETSc 3.21.5 (https://petsc.org/release/)
\item METIS 5.1.0 (https://github.com/KarypisLab/METIS)
\end{itemize}

On the different clusters, we also use some additional software packages, which are provided by their respective \emph{module }environments. They can be seen in Table~\ref{tab:system-configuration}, software row.

\artinpart{Datasets / Inputs}

Input matrices can be downloaded from the SuiteSparse collection at https://sparse.tamu.edu/, in the Matrix Market file format.  The following six matrices from SuiteSparse are used: Bump\_2911, Geo\_1438, Flan\_1565, Queen\_4147, Serena and audikw\_1.  For each matrix, partitionings are included in artifact $A_1$.

\artinpart{Installation and Deployment}

To install the program the following minimum software dependencies must be satisfied:
\begin{itemize}
\item CMake 3.12 or newer
\item A C/C++ compiler compatible with C++-17
\item For NVIDIA GPUs: NVIDIA CUDA Compiler (NVCC), cuBLAS and cuSPARSE from CUDA Toolkit 11.6 or newer
\end{itemize}

The following optional software packages are needed to enable some
features:
\begin{itemize}
\item METIS 5.1.0 is required for multi-GPU matrix partitioning, unless using precomputed partitions (e.g., from artifact $A_1$).
\item NVIDIA Collective Communications Library (NCCL) version 2.18.5 or newer is needed to use NCCL-based communication for NVIDIA GPUs
\item PETSc 3.17 or newer with CUDA support enabled is needed to use PETSc's PIPE-PCG solver
\item zlib is needed to use gzip-compressed Matrix Market files as input
\item If the compiler supports OpenMP, then it can be enabled to use multiple threads to speed up preprocessing steps.
\end{itemize}

\artcomp

The experimental workflow consists of a single task: running the compiled executable three times for each matrix, solver, and GPU configuration. The program outputs detailed metrics, such as runtime and iteration counts. For further details and the job submission scripts used on the three clusters, please refer to artifact $A_1$

\artout

To analyze the experimental outputs, we extract the solver runtimes from the log files and calculate the average across the three trials for each configuration of matrix, solver, and GPU count. Using these average runtimes, we compute the performance speedup for $N$ GPUs by dividing the baseline two-stream runtime by the runtime of the three-stream implementation for that specific matrix and GPU allocation.

\appendixAE

\arteval{1}
\artin

All commands are run from the root directory of the artifact,
\cmd{\$\{artifact\_root\}}, on a login node of the cluster.  Below,
\cmd{\$\{system\}} is \cmd{marenostrum5} or \cmd{leonardo}.  Each
system directory \cmd{scripts/\$\{system\}/} contains a README with
further details.

\artinpart{Environment setup}

METIS 5.1.0 and PETSc 3.21.5 (with 32-bit indices) must
be installed in a directory given by \cmd{ACG\_PKGS\_DIR} (default:
\cmd{\$\{HOME\}/pkgs}), in the subdirectories \cmd{metis-32-5.1.0} and \cmd{petsc-32-3.21.5}; on Leonardo,
\cmd{glibc-2.29} is also needed.  The module files for these packages
are included in \cmd{scripts/\$\{system\}/modulefiles} and are loaded
from there.  The Slurm account is given through environment variables:

\cmd{\$ export ACG\_PKGS\_DIR=/path/to/pkgs}

\cmd{\$ export SBATCH\_ACCOUNT=<project>}

\cmd{\$ export SBATCH\_QOS=acc\_ehpc}

\noindent
where the QOS is needed on MareNostrum~5 only and depends on the type
of allocation.  The environment is then set up by sourcing the script for the system,
which loads the modules and creates the directory \cmd{logs/} for the
Slurm output:

\cmd{\$ . scripts/\$\{system\}/envsetup-\$\{system\}.sh}

\artinpart{aCG Library Install}

aCG is built with the build script for the system:

\cmd{\$ scripts/\$\{system\}/build-acg-cuda.sh}

\noindent
The script copies \cmd{src/aCG} to \cmd{build/\$\{system\}/src},
applies \cmd{scripts/\$\{system\}/\$\{system\}.patch} (which disables
NCCL CollNet for the split communicators, and on Leonardo also the
internal profiling, as in the binaries used for the paper), configures
the CUDA version of aCG with CMake and compiles it.  The executable is
installed as \cmd{bin/acg-cuda}.  On MareNostrum~5, the configuration
uses \cmd{nvc}, CUDA 12.9 \cmd{nvcc} and an empty
\cmd{CMAKE\_BUILD\_TYPE}; on Leonardo, it uses \cmd{nvc}, CUDA 12.3
\cmd{nvcc}, \cmd{CMAKE\_BUILD\_TYPE=release} and
\cmd{CMAKE\_CUDA\_ARCHITECTURES=80}.

\artinpart{Input matrices and partitions}

The six SuiteSparse matrices (\texttt{audikw\_1}, \texttt{Bump\_2911},
\texttt{Flan\_1565}, \texttt{Geo\_1438}, \texttt{Queen\_4147} and
\texttt{Serena}) are not included in the artifact because of their size;
only their METIS row partitions are. Before running any job, download the
matrices in Matrix Market format from the SuiteSparse Matrix
Collection~\cite{davis2011university} by running, from the artifact root
directory:

\cmd{scripts/download-matrices.sh}

The script downloads each matrix from
\url{https://sparse.tamu.edu/MM/<group>/<matrix>.tar.gz}, extracts it to
\texttt{data/suitesparse/mtx/<matrix>.mtx}, decompresses the partitions in
\texttt{data/suitesparse/partitions/}, and verifies all files against the
checksums in \texttt{data/SHA256SUMS}. It requires \texttt{curl} and about
3.5\,GB of disk space for the downloads and 11\,GB for the extracted
matrices; the downloaded archives are deleted after extraction unless
\texttt{--keep-tarballs} is given. Individual matrices can be downloaded
by naming them, e.g.\ \texttt{scripts/download-matrices.sh Queen\_4147}.
Re-running the script skips matrices that are already present.

The matrix poisson3da\_12 and its partitions are
obtained by downloading and unpacking $A_2$ and placing the files at:

\cmd{data/poisson3da/mtxbin32/}

\cmd{\quad poisson3da\_12\_A.mtxbin32.gz}

\cmd{data/poisson3da/partitions/}

\cmd{\quad poisson3da\_12\_A\_rowpart\_metis\_}

\cmd{\quad <num\_of\_part>\_parts.mtx.gz}

\noindent


\artcomp

Each experiment is a Slurm job script that runs the six solver
variants, three times each, for one matrix and one number of nodes.
The job scripts are located in

\cmd{scripts/\$\{system\}/suitesparse/\$\{matrix\}/}

\cmd{scripts/\$\{system\}/poisson3da/poisson3da\_12/}

\noindent
and are named

\cmd{\$\{matrix\}-rtol-0-\$\{nodes\}-nodes-}

\cmd{\quad \$\{gpus\}-procs.sh}

\noindent
where \cmd{\$\{nodes\}} and \cmd{\$\{gpus\}} are zero-padded to three
and four digits.  For example, Queen\_4147 on 4 nodes (16 GPUs) of Leonardo is run with:

\cmd{\$ sbatch scripts/leonardo/suitesparse/}

\cmd{\quad Queen\_4147/}

\cmd{\quad Queen\_4147-rtol-0-004-nodes-}

\cmd{\quad 0016-procs.sh}

\noindent
There are 30 SuiteSparse jobs per system (6 matrices on 1, 2, 4, 8
and 16 nodes) and 3 (MareNostrum~5: 16, 32 and 64 nodes) or 4
(Leonardo: 16, 32, 64 and 128 nodes) jobs for poisson3da\_12.  All jobs
of a system can be submitted with:

\cmd{\$ for j in scripts/\$\{system\}/*/*/*.sh; do}

\cmd{\quad sbatch \$j; done}

\noindent
Within the 8-hour evaluation budget, we suggest running the 30
SuiteSparse jobs on one system (about 30 node hours) and, if the
allocation permits, the 16-node poisson3da\_12 job.

The output of each run is written to

\cmd{results/acg/\$\{date\}/\$\{variant\}/\$\{matrix\}/}

\noindent
A run that
exits with a non-zero status keeps the suffix \cmd{.tmp}; this is
expected for the PCG variants, which stop at the iteration limit, and
these files are used in the analysis.

\artout

The solver times of all runs of a given date are collected into a CSV
file, with one row per variant, matrix and number of GPUs and one
column per run:

\cmd{\$ python3 scripts/collect/collect\_results.py}

\cmd{\quad results/acg/\$\{date\} H100 > results.csv}

\noindent
(\cmd{A100} on Leonardo).  From this file, the figures of the paper
are produced with Python 3, pandas and matplotlib.  The script
\cmd{figure2x3.py} plots the six SuiteSparse matrices and
\cmd{figure1x3\_large\_matrices.py} the large matrices, for the solver
family given by \cmd{--solver}, for example:

\cmd{\$ python3 scripts/figures/figure2x3.py}

\cmd{\quad results.csv --solver pcg}

\cmd{\quad --system MareNostrum5 -o ss\_pcg}

\noindent
which writes \cmd{ss\_pcg.png} and \cmd{ss\_pcg.pdf}.  Each bar is
the mean of the three runs, with the standard deviation as error bar,
and the speedups of the three-stream solver are printed under each
panel.  The variants \cmd{PIPE-*-PETSC}, \cmd{PIPE-*-nccl} and
\cmd{PIPE-*-nccl-split} correspond to ``PETSc'', ``Two-Stream'' and
``Three-Stream'' in the paper.  For the SuiteSparse matrices, the
resulting figures correspond to Figure~\ref{fig:MN5_suitesparse} and Figure~\ref{fig:MN5_suitesparse_bicgstab}
(MareNostrum~5); for poisson3da\_12, they
correspond to the poisson3da\_12 panels of Figure~\ref{fig:Leonardo_large_matrices}, Figure~\ref{fig:MN5_large_matrices_bicgstab}.The three-stream solvers are expected to have the
lowest solver times, with speedups over the two-stream and PETSc
solvers similar to those reported in the paper.


\subsection*{References}
{\small
Davis, T. A. and Y. Hu. 2011. The University of Florida Sparse Matrix Collection. {\itshape ACM Trans Math Softw}. 38, 1.

\noindent Karypis, G., and V. Kumar. 1998. A fast and high quality multilevel scheme for partitioning irregular graphs. {\itshape SIAM J Sci Comput}. 20, 1.}